# A SUPER-CONDUCTING LINAC INJECTOR FOR THE BNL-AGS[*]

D. Raparia, A.G. Ruggiero, BNL, Po Box 5000, Upton, NY 11973, USA

*Abstract*

This paper reports on the feasibility study of a proton Super-Conducting Linac (SCL) as a new injector to the Alternating Gradient Synchrotron (AGS) of the Brookhaven National Laboratory (BNL). The Linac beam energy is in the range of 1.5 to 2.4 GeV. The beam intensity is adjusted to provide an average beam power of 4 MW at the top energy of 24 GeV. The repetition rate of the SCL-AGS facility is 5 beam pulses per second.

## 1 INTRODUCTION

It has been proposed to upgrade the AGS accelerator complex to provide an average proton beam power of 4 MW at the energy of 24 GeV. The facility can be used either as a proton driver for the production of intense muon and neutrino beams, or/and as a pulsed high-energy spallation neutron source. The upgrade requires operation of the accelerator at the rate of five cycles per second, and a new injector for an increase of the AGS beam intensity by a factor of three.

In a separate technical report [1] we have described the methods and the requirements to operate the BNL-AGS accelerator facility at the rate of 5 proton beam pulses per second. The major requirement is an extensive addition and modification of the present power supply system. At the same time, a new injector to the AGS, capable to operate at the rate of 5 beam pulses per second, and to provide a three-fold increase of the present beam intensity, is required. The present injector made of the 200-MeV Linac and of the 1.5-GeV AGS-Booster will not be able to fulfill the goals of the upgrade.

The proposed new injector is a 1.5 - 2.4 GeV SCL with an average output beam power of 250 - 400 kW. The largest energy is determined by the need to control beam losses due to stripping of the negative ions that are used for multi-turn injection into the synchrotron. The high-energy case is to be preferred if one wants to reduce the effects of space charge, for a sufficiently small beam transverse size that fits the AGS acceptance. The duty cycle is about a half percent.

The paper describes the preliminary design of the SCL. It is composed of three parts; a front end, that is a 60-mA negative-ion source, followed by a 5-MeV RFQ, a room temperature Drift-Tube Linac (DTL) that accelerates protons to 150 MeV, and the SCL proper. This in turn is made of four sections, each with its own energy range, and different cavity-cryostat arrangement. The four sections are labeled: Low-Energy (LE), Medium Low-Energy (MLE), Medium High-Energy (MHE), and High-Energy (HE).

## 2 THE NEW INJECTOR

The two different Linac energy cases are compared in Table 1. AGS performance during multi-turn injection is also summarized in the same Table. Both cases correspond to the same average beam current of 0.17 mA, that yields the same average beam power of 4 MW at the top energy of 24 GeV. The repetition rate of 5 beam pulses per second is assumed in both cases; that gives the same intensity of $2.1 \times 10^{14}$ protons accelerated per AGS cycle; that is a factor of 3 higher then the intensity routinely obtained with the present injector. At the end of injection, that takes about 317-330 turns, the space-charge tune depression is $\Delta\nu = 0.4$ at the energy of 1.5 GeV, and $\Delta\nu = 0.2$ at 2.4 GeV, assuming a bunching factor (the ratio of beam peak current to average current), during the early part of the acceleration cycle, of 4. Also, with the same normalized beam emittance of 250 $\pi$ mm-mrad, the actual beam emittance is smaller at 2.4 GeV, $\varepsilon = 73$ $\pi$ mm-mrad, versus 104 $\pi$ mm-mrad at 1.5 GeV. Obviously, the effective acceptance of the AGS at injection is to be larger than these beam emittance values.

The front-end of the Linac is made of an ion source operating with a 1% duty cycle at the repetition rate of 5 pulses per second. The beam current within a pulse is 60 mA of negative-hydrogen ions. The ion source seats on a platform at 35-60 kVolt, and is followed by a 5-MeV RFQ that works at 201.25 MHz. The beam is pre-chopped by a chopper located between the ion source and the RFQ. The beam chopping extends over 75% of the beam length, at a frequency matching the accelerating rf (4.11-4.28 MHz) at injection into the AGS. Moreover, the transmission efficiency through the RFQ is taken conservatively to be 80%, so that the average current of the beam pulse in the Linac, where we assume no further beam loss, is 36 mA.

The combination of the chopper and of the RFQ pre-bunches the beam with a sufficiently small longitudinal extension so that each of the beam bunches can be entirely fitted in the accelerating rf buckets of the following DTL that operates at either 402.5 or 805 MHz. The DTL is a room temperature conventional Linac that accelerates to 150 MeV.

---

Table 1. Injector and AGS Parameters for the Upgrade

| Linac Aver. Power, MW | 0.25 | 0.40 |
|---|---|---|
| Kinetic Energy, GeV | 1.5 | 2.4 |
| β | 0.9230 | 0.9597 |
| Momentum, GeV/c | 2.2505 | 3.2037 |
| Magnetic Rigidity, T-m | 7.5068 | 10.6862 |
| AGS Circumference, m | 807.12 | 807.12 |
| Revol. Frequency, MHz | 0.3428 | 0.3565 |
| Revolution Period, μs | 2.9169 | 2.8053 |
| Bending Radius | 79.832 | 79.832 |
| Injection Field | 0.9403 | 1.3386 |
| Protons per Turn, x $10^{11}$ | 6.563 | 6.312 |
| Number of injected Turns | 317 | 330 |
| Beam Pulse Length, ms | 0.9259 | 0.9259 |
| Duty Cycle, % | 0.4630 | 0.4630 |
| Emittance, π mm-mrad | 104. | 73. |
| Space-Charge Δν | 0.41 | 0.21 |

## 3 THE SUPER-CONDUCTING LINAC

The SCL accelerates the proton beam from 150 MeV to 1.5 or 2.4 GeV. The configuration and the design procedure of the SCL is described in detail in [2]. It is typically a sequence of a number of identical periods as shown in Figure 1. Each period is made of a cryo-module of length $L_{cryo}$ and of an insertion of length $L_{ins}$. The insertion is needed for the placement of focusing quadrupoles, vacuum pumps, steering magnets, beam diagnostic devices, bellows and flanges. It can be either at room temperature or in a cryostat as well. Here we assume that the insertions are at room temperature. The cryo-module includes $M$ identical cavities, each of $N$ identical cells, and each having a length $NL_{cell}$, where $L_{cell}$ is the length of a cell. Cavities are separated from each other by a drift space $d$. An extra drift of length $L_w$ may be added internally on both sides of the cryo-module to provide a transition between cold and warm regions. Thus,

$$L_{cryo} = MN L_{cell} + (M-1) d + 2 L_w \qquad (1)$$

There are two symmetric intervals: a minor one, between the two middle points A and B, as shown in Figure 1, that is the interval of a cavity of length $NL_{cell} + d$; and a major one, between the two middle points C and D, that defines the range of a period of total length $L_{cryo} + L_{ins}$. Thus, the topology of a period can be represented as a drift of length $g$, followed by $M$ cavity intervals, and a final drift of length $g$, where

$$g = L_w + (L_{ins} - d)/2 \qquad (2)$$

The choice of cryo-modules with identical geometry, and with the same cavity/cell configuration, is economical and convenient for construction. But there is, nonetheless, a penalty due to the reduced transit-time-factors when a particle crosses cavity cells, with length adjusted to a common central value $β_0$ that does not correspond to the particle instantaneous velocity. To minimize this effect the SCL is divided in four sections, each designed around a different central value $β_0$, and, therefore, with different cavity/cell configuration. The cell length in a section is fixed to be

$$L_{cell} = λβ_0/2 \qquad (3)$$

where $λ$ is the rf wavelength. We assume the same operating frequency of 805 MHz for the entire SCL, so that $λ = 37.24$ cm.

The major parameters of the four sections of the SCL are given in Tables 2 and 4. Transverse focussing is done with a sequence of FODO cells with half-length equal to that of a period. The phase advance per cell is 90°. The rms normalized betatron emittance is 0.3 π mm-mrad. The rms bunch area is 0.5 π °MeV. The rf phase angle is 30°. The cost estimate (with no contingency) for each section of the SCL has been made assuming the cost and rf parameters shown in Table 3. The total expected cost is around 300 M$, including also the front-end and the room-temperature DTL.

The length of the Linac depends on the average accelerating gradient. The local gradient has a maximum value that is limited by three causes: (1) The surface field limit at the frequency of the 805 MHz is 26 MV/m. For a realistic cavity shape, we set a limit of a 13 MV/m on the axial electric field. (2) There is a limit on the power provided by rf couplers that we take here not to exceed 400 kW, including a contingency of 50% to avoid saturation effects. (3) To make the longitudinal motion stable, we can only apply an energy gain per cryo-module that is a relatively small fraction of the beam energy in exit of the cryo-module. The conditions for stability of motion have been derived in [2].

The proposed mode of operation is to operate each section of the SCL with the same rf input power per cryo-module. This will result to some variation of the actual axial field from one cryo-module to the next. If one requires also a constant value of the axial field, this may be obtained by adjusting locally the value of the rf phase.

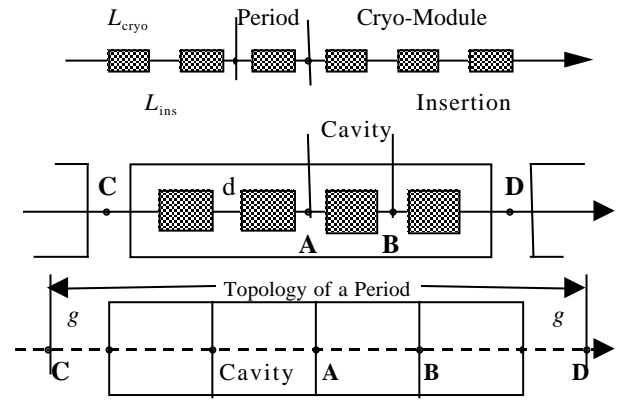

Figure 1: Configuration of a Proton SCL Accelerator.

The number of cells and cavities varies from insertion to insertion. The number of couplers varies for 1 to 2. The total length of the injector including the front-end and the DTL is expected to be about 500 meters.

It is proposed to build the entire SCL in two stages. During the first stage the final energy if 1.5 GeV, and the Linac is made of the three first sections. In a second stage the High-Energy section is added for the final energy of 2.4 GeV, if indeed this should result to be necessary.

Table 2. General Parameters of the SCL

| Linac Section | LE | MLE | MHE | HE |
|---|---|---|---|---|
| Av. Beam Power, kW | 50 | 133 | 250 | 400 |
| Av. Beam Current, mA | 0.167 | 0.167 | 0.167 | 0.167 |
| In. Kin. Energy, GeV | 0.150 | 0.300 | 0.800 | 1.500 |
| Fin. Kin. Energy, GeV | 0.300 | 0.800 | 1.500 | 2.400 |
| Frequency, MHz | 805 | 805 | 805 | 805 |
| Protons / Bunch x $10^9$ | 1.49 | 1.49 | 1.49 | 1.49 |
| Temperature, °K | 2.0 | 2.0 | 2.0 | 2.0 |
| Cells / Cavity | 4 | 4 | 4 | 4 |
| Cavities / Cryo-Module | 4 | 8 | 8 | 8 |
| Cavity Separation, cm | 32 | 32 | 32 | 32 |
| Cold-Warm Trans., cm | 30 | 30 | 30 | 30 |
| Cavity Int. Diam., cm | 10 | 10 | 10 | 10 |
| Length of Warm Ins., m | 1.079 | 1.079 | 1.079 | 1.079 |
| Acc. Gradient, MeV/m | 12.2 | 11.9 | 12.9 | 12.4 |
| Cavities / Klystron | 4 | 8 | 4 | 4 |
| rf Couplers / Cavity | 1 | 1 | 2 | 2 |

Negative ion stripping during transport down the SCL has been found to be very negligible. But a final 30° bend, before injection into the AGS, could be of a concern [3]. To control the rate of beam loss by stripping to a $10^{-4}$ level, the bending field should not exceed 2.6 kGauss over a total integrated bending length of 15 m, in the 1.5 GeV case. At 2.4 GeV, the bending field is 1.9 kGauss, and the integrated bending length about 30 m.

Table 3. Cost ('00 $) and Other Parameters

| AC-to-rf Efficiency | 0.45 | For pulsed mode |
|---|---|---|
| Cryo. Efficiency | 0.004 | At 2.0 °K |
| Electricity Cost | 0.05 | $ / kWh |
| Linac Availability | 75 | % of yearly time |
| Normal Cond. Cost | 150 | k$ / m |
| Superconducting Cost | 500 | k$ / m |
| Tunnel Cost | 100 | k$ / m |
| Cost of Klystron | 2.50 | $ / W of rf Power |
| Cost of Refrig. Plant | 2 | k$ / W @ 2.0 °K |
| Cost of Elect. Distrib. | 0.14 | $ / W of AC Power |

There are two problems in the case of the pulsed-mode of operation of the SCL. First, there are Lorentz forces that deform the cavity cells out of resonance. They can be controlled with a thick cavity wall strengthened to the outside by supports. Second, there is an appreciable period of time to fill the cavities with rf power before the maximum gradient is reached [2]. During the filling time, extra power is dissipated also before the beam is injected into the Linac. The extra amount of power is the ratio of the filling time to the beam pulse length.

Table 4. Summary of the SCL Design

| Linac Section | LE | MLE | MHE | HE |
|---|---|---|---|---|
| Energy: in | 0.15 | 0.30 | 0.80 | 1.50 |
| out, GeV | 0.30 | 0.80 | 1.50 | 2.40 |
| Velocity, β: in | 0.5066 | 0.6526 | 0.8418 | 0.9230 |
| out | 0.6526 | 0.8418 | 0.9230 | 0.9597 |
| **Cell Reference $\beta_0$** | **0.530** | **0.680** | **0.850** | **0.935** |
| Cell Length, cm | 9.87 | 12.66 | 15.83 | 17.41 |
| Total No. of Periods | 9 | 12 | 13 | 15 |
| Length of a period, m | 4.218 | 7.971 | 8.984 | 9.490 |
| FODO ampl. func., $\beta_Q$, m | 14.40 | 27.21 | 30.67 | 32.40 |
| *Total Length, m* | *37.96* | *95.65* | *116.79* | *142.35* |
| Coupler rf Power, kW (*) | 300 | 375 | 255 | 270 |
| En. Gain/Period, MeV | 16.67 | 41.67 | 56.67 | 60.00 |
| Total No. of Klystrons | 9 | 12 | 26 | 30 |
| Klystron Power, kW (*) | 1200 | 3000 | 2040 | 2160 |
| Cavity Filling Time, ms | 0.30 | 0.18 | 0.12 | 0.12 |
| $Z_0T_0^2$, ohm/m | 271.8 | 447.4 | 699.0 | 845.8 |
| $Q_0$ x $10^9$ | 5.5 | 7.0 | 8.7 | 9.6 |
| Ave. Dissip. Power, kW | 0.009 | 0.012 | 0.009 | 0.008 |
| Ave. HOM-Power, kW | 0.0016 | 0.0042 | 0.0046 | 0.0053 |
| Ave. Cryog. Power, kW | 0.152 | 0.430 | 0.527 | 0.644 |
| Ave. Beam Power, MW | 0.025 | 0.083 | 0.117 | 0.150 |
| Ave. rf Power, MW (*) | 0.050 | 0.149 | 0.197 | 0.248 |
| AC Power for rf, MW (*) | 0.110 | 0.331 | 0.438 | 0.552 |
| AC Power for Cryo., MW | 0.038 | 0.107 | 0.132 | 0.161 |
| **AC Power, MW (*)** | **0.148** | **0.439** | **0.570** | **0.713** |
| **Efficiency, % (*)** | **16.9** | **19.0** | **20.5** | **21.0** |
| Capital Cost '00 M$: | | | | |
| Rf Klystron (*) | 0.124 | 0.373 | 0.493 | 0.621 |
| Electr. Distr. (*) | 0.021 | 0.061 | 0.080 | 0.100 |
| Refrig. Plant | 0.303 | 0.860 | 1.055 | 1.288 |
| Warm Structure | 1.619 | 2.104 | 2.266 | 2.590 |
| Cold Structure | 14.126 | 41.351 | 51.381 | 63.085 |
| Tunnel | 3.904 | 9.673 | 11.787 | 14.343 |
| **Cost, '00 M$ (*)** | **20.096** | **54.422** | **67.061** | **82.027** |
| **Operat, '00 M$/y (*)** | **0.049** | **0.144** | **0.187** | **0.234** |

(*) Including 50% rf power contingency.